\documentclass[3p]{elsarticle}
\usepackage{color}
\usepackage{array}
\usepackage{bm}
\usepackage{etex}
\usepackage{multirow}
\usepackage{amsmath}
\usepackage{amssymb}
\usepackage{graphicx}
\usepackage{wasysym}
\usepackage{tikz}
\usepackage{forloop}
\usepackage{graphicx}
\usepackage{bm}
\usepackage{soul}
\usepackage{color,xcolor}
\usepackage{amsfonts}
\usepackage{mathrsfs,amssymb}
\usepackage{physics}
\usepackage{pifont}
\usepackage{xcolor}
\usepackage{amsmath}
\usepackage{multirow}
\usepackage[caption=false]{subfig}
\usepackage{graphicx}
\usepackage{float}
\usepackage{tabularx}
\usepackage{rotating}
\usepackage{tcolorbox}
\usepackage[unicode=true,
bookmarks=true,bookmarksnumbered=false,bookmarksopen=false,
breaklinks=false,pdfborder={0 0 0},pdfborderstyle={},backref=false,colorlinks=true]
{hyperref}

\makeatletter

\biboptions{sort&compress}

\usepackage{xcolor}
\usepackage{textcomp}
\usepackage{hyphenat}
\usepackage{makecell} 
\usepackage[]{lmodern} 
\usepackage{pifont}
\usepackage{listings}

\makeatother

\usepackage{listings}

\usepackage[normalem]{ulem}

\newcommand{\delete}[1]{{\color{blue}\sout{#1}}}

\begin{document}
	
	\title{\texttt{ToMSGKpoint}: A user-friendly package for computing symmetry transformation properties of electronic eigenstates of nonmagnetic and magnetic crystalline materials}
	
	\author[bit1,bit2]{Liangliang Huang}
	
	\cortext[cor]{Corresponding author}
	
	\author[bit1,bit2]{Xiangang Wan}
	
	\author[bit1,bit2]{Feng Tang\corref{cor}}

	\ead{fengtang@nju.edu.cn}
	
	\address[bit1]{National Laboratory of Solid State Microstructures and School of Physics, Nanjing University, Nanjing 210093}
	
	\address[bit2]{Collaborative Innovation Center of Advanced Microstructures, Nanjing University, Nanjing 210093, China}
	
	\begin{abstract}
The calculation of (co)irreducible representations of energy bands at high-symmetry points (HSPs) is essential for high-throughput research on topological materials based on symmetry-indicators or topological quantum chemistry. However, existing computational packages usually require transforming crystal structures into specific conventions, thus hindering extensive application, especially to materials whose symmetries are yet to be identified. To address this issue, we developed a Mathematica package, \texttt{ToMSGKpoint}, capable of determining the little groups and (co)irreducible representations of little groups of HSPs, high-symmetry lines (HSLs), and high-symmetry planes (HSPLs) for any nonmagnetic and magnetic crystalline materials in two and three dimensions, with or without considering spin-orbit coupling. To the best of our knowledge, this is the first package to achieve such functionality. The package also provides magnetic space group operations, supports the analysis of (co)irreducible representations of energy bands at HSPs, HSLs, and HSPLs using electronic wavefunctions obtained from \textit{ab initio} calculations interfaced with VASP. Designed for user convenience, the package generates results in a few simple steps and presents all relevant information in clear tabular format. Its versatility is demonstrated through applications to nonmagnetic topological insulator Bi$_2$Se$_3$ and Dirac semimetal Na$_3$Bi, as well as the antiferromagnetic topological material MnBi$_2$Te$_4$. Suitable for any crystal structure, this package can be conveniently  applied in a streamlined study once magnetic space group varies with various symmetry-breakings caused by phase transitions.
	\end{abstract}
	\maketitle	
	
	\section*{Program summary}
	
	\noindent \textit{Program title}: \texttt{ToMSGKpoint}
	
	\noindent \textit{Developer's respository link}: \url{https://github.com/FengTang1990/ToMSGKpoint}
	
	\noindent\textit{Licensing provisions}: GNU General Public Licence
	3.0
	
	\noindent\textit{Distribution format}: zip
	
	\noindent\textit{Programming language}: Wolfram
	
	\noindent\textit{Nature of problem}: The package \texttt{ToMSGKpoint} provides magnetic space group operations for any crystal structure, along with the little groups of high-symmetry points, lines, and planes, and their corresponding (co)irreducible representations. It also facilitates the transformation from a customized primitive crystal structure to the Bradley-Cracknell convention. Furthermore, based on electronic wavefunctions obtained from VASP calculations, the package computes the (co)irreducible representations of energy bands at high-symmetry points, lines, and planes.
	
	\noindent\textit{Solution method}: We do not directly calculate the (co)irreducible representations of the little groups at high-symmetry points, lines, and planes. Instead, we first determine the transformation from the given crystal structure to the Bradley-Cracknell convention. Using this transformation, we derive the (co)irreducible representations for the little groups in the customized primitive crystal structure convention based on those in the Bradley-Cracknell convention.

\section{Introduction}\label{1}
In recent years, the topological properties of materials have garnered significant attention, with high-throughput classification of topological materials playing a crucial role in this field. Over the past decade, substantial progress has been made in the high-throughput classification of nonmagnetic \cite{1,2,3}, magnetic \cite{6}, and phononic materials \cite{7}, driven by advances in topological quantum chemistry \cite{4} or symmetry indicator theory \cite{5}. The cornerstone of both topological quantum chemistry and symmetry indicator theory lies in calculating the irreducible representations (irreps) or coirreducible representations (coirreps) of the band structure at high-symmetry points (HSPs). The little groups at HSPs, high-symmetry lines (HSLs), or high-symmetry planes (HSPLs), along with their (co)irreps, play an essential role not only in the topological classification of materials but also in constructing $k\cdot p$ models for experimental and theoretical studies \cite{k1,k2,k3,k4}. They are also instrumental in understanding the nature of band degeneracies responsible for nontrivial physical phenomena.

The labeling of (co)irreps for little groups at $\vec{k}$-points in the same space group or magnetic space group (MSG) can vary depending on the conventions used for the crystal structure. The most commonly adopted convention aligns with the Bilbao Crystallographic Server (BCS). Several packages based on the BCS convention exist for calculating irreps of energy bands in nonmagnetic materials, including \texttt{Irvsp} \cite{c1}, \texttt{SpaceGroupIrep} \cite{c2} (indirectly based on the BCS convention via the \texttt{vasp2trace} tool, which itself adheres to the BCS convention), \texttt{IrRep} \cite{c3}, and \texttt{qeirreps} \cite{c4}. However, for magnetic space groups, there is currently no package that directly calculates (co)irreps of energy bands at arbitrary  $\vec{k}$-points. Although \texttt{MagVasp2trace} \cite{6}, based on the BCS convention, can generate \texttt{trace.txt} files to be used by the "Check Topological Magnetic Mat." tool on the BCS website to determine the topological properties of magnetic materials, it does not directly output the (co)irreps of energy bands at arbitrary  $\vec{k}$-points. \texttt{MSGCorep} \cite{c5} extends the functionality by calculating irreps of energy bands at arbitrary  $\vec{k}$-points based on \texttt{trace.txt} files generated by \texttt{MagVasp2trace}. However, it still has significant limitations: users must transform crystal structures into the BCS convention, and for systems with spin-orbit coupling (SOC), modifications to the original \texttt{MagVasp2trace} package are necessary to generate the correct \texttt{trace.txt} file for use with \texttt{MSGCorep}.

To address these challenges, we have developed a Mathematica-based computational package, \texttt{ToMSGKpoint}, which overcomes these difficulties and provides the following key functionalities:
\begin{enumerate}
	\item Determines the MSG type of any given nonmagnetic or magnetic crystalline material, along with its MSG operations (see the \texttt{GetMSGOP} function in Sec.~\ref{3.1}).
	\item Provides the transformation from the convention of the customized primitive crystal structure to the Bradley-Cracknell (BC) convention (see the \texttt{GetMSG} function in Sec.~\ref{3.1}).
	\item Computes the little groups and their (co)irreps at HSPs, HSLs, and HSPLs (see the \texttt{ToHSPInfo}, \texttt{ToHSLInfo}, and \texttt{ToHSPLInfo} functions in Sec.~\ref{3.1}).
	\item For systems with and without SOC, computes the (co)irreps of any band at HSPs, HSLs, and HSPLs based on electronic wavefunctions obtained from the \textit{ab initio} package VASP \cite{v1,v2}. Additionally, it calculates the distribution of valence band (co)irreps across all HSPs and generates input files for CalTopoEvol \cite{tang} to compute the topological properties of materials(see the \texttt{GetKpointTop} function in Sec.~\ref{3.1}).
\end{enumerate}

To the best of our knowledge, this is the first package to achieve such functionality. Moreover, \texttt{ToMSGKpoint} is also compatible with two-dimensional materials, which are experimentally tunable, offering significant convenience for experimental research.

It should be noted that our approach differs fundamentally from conventional methods. Traditional approaches typically require transforming the crystal structure into some specific convention and determining the little groups and their (co)irreps at HSPs, HSLs, and HSPLs. Then, based on first-principles calculated wavefunctions, the (co)irreps of the bands can be computed. In contrast, our approach eliminates the need to adhere to a particular convention. Instead, we directly compute the MSG operations from the given crystal structure, determine the coordinates of HSPs, HSLs, and HSPLs, and identify their corresponding little groups and (co)irreps. From this, we proceed with material calculations. Since we already have the data for the little groups and their (co)irreps under the BC convention at HSPs, HSLs, and HSPLs, we do not directly calculate the (co)irreps of the little groups at these points. Instead, our method first determines the transformation between the user’s structural convention and the BC convention. This transformation is then used to indirectly derive the (co)irreps of the little groups at HSPs, HSLs, and HSPLs.

The structure of this paper is as follows: In Sec. \ref{2}, we present the transformation relations from any given primitive crystal structure convention to the BC convention and explain how to calculate the representation matrix of a unitary MSG operation of the little groups at any $\vec{k}$-point based on VASP-calculated electronic wavefunctions, as well as the determination of the (co)irreps of the band structure at the $\vec{k}$-point. In Sec. \ref{3}, we introduce the installation and usage of the package \texttt{ToMSGKpoint}. In Sec. \ref{4}, we demonstrate the application of the package \texttt{ToMSGKpoint} using nonmagnetic topological insulator Bi$_2$Se$_3$ and Dirac semimetal Na$_3$Bi, as well as the antiferromagnetic topological material MnBi$_2$Te$_4$, with SOC included. Finally, in Sec. \ref{5}, we provide the conclusion and perspective.

\section{Theory}\label{2}
First, let us discuss the transformation relations from any customized primitive crystal structure convention to the BC convention.

For any given crystal structure, assume that its primitive unit cell lattice vectors are $(\vec{a}_1, \vec{a}_2, \vec{a}_3)$ and the Cartesian coordinate system is $Oxyz$. We use three parameters $(\vec{\delta}, P, U_{xyz})$ to represent the transformation from the customized primitive crystal structure convention to the BC convention. The transformation from $(\vec{a}_1, \vec{a}_2, \vec{a}_3)$ to the conventional lattice vectors $(\vec{a}, \vec{b}, \vec{c})$ in the BC convention is:
\begin{equation}
	(\vec{a}, \vec{b}, \vec{c}) = (\vec{a}_1, \vec{a}_2, \vec{a}_3)\cdot P.
\end{equation}

Likewise, the transformation from $Oxyz$ to the Cartesian coordinate system $O'x'y'z'$ in the BC convention is:
\begin{equation}
	(x', y', z') = P \cdot (x, y, z) + \vec{\delta},
\end{equation}
where the new coordinates $(x', y', z')$ are obtained by applying the transformation matrix $P$ to the original coordinates $(x, y, z)$, followed by a translation given by $\vec{\delta}$ ($\vec{\delta} = \delta_1 \vec{a}_1 + \delta_2 \vec{a}_2 + \delta_3\vec{a}_3$). The transformation from the unit vectors $ \left( \vec{e}_{x} , \vec{e}_{y} , \vec{e}_{z} \right) $ of $Oxyz$ to the unit vectors $ \left( \vec{e}_{x'} , \vec{e}_{y'} , \vec{e}_{z'} \right) $ of $O'x'y'z'$ is expressed as:
\begin{equation}
	\left( \vec{e}_{x'} , \vec{e}_{y'} , \vec{e}_{z'} \right) = \left( \vec{e}_{x} , \vec{e}_{y} , \vec{e}_{z} \right) \cdot U_{xyz}.
\end{equation}

After discussing the transformation relations, we now address the calculation of the representation matrix for a unitary MSG operation of the little group at any $\vec{k}$-point based on VASP calculations, as well as the determination of the (co)irreps of the band structure at the $\vec{k}$-point. Notably, unitary operations of the little groups suffice to determine the (co)irreps of the energy band structure. In band theory, the Schr\"odinger equation is given by:  
	\begin{equation}
		H \psi_{n\vec{k}}(\vec{r}) = E_{n\vec{k}} \psi_{n\vec{k}}(\vec{r}),
	\end{equation}  
	where \( H \) is the Hamiltonian operator, which includes the kinetic energy, the periodic potential \( V(\vec{r}) \), and, when considering SOC, the SOC term \( H_{\text{SOC}} \). The wavefunction \( \psi_{n\vec{k}}(\vec{r}) \) represents the Bloch wavefunction for the \( n \)-th energy band at wavevector \( \vec{k} \), and \( E_{n\vec{k}} \) is the corresponding energy eigenvalue. In the context of VASP, the Bloch wavefunction is expanded in a plane wave (PW) basis. When considering SOC, the basis is extended to include both spin states, specifically $\lbrace \ket{\uparrow}, \ket{\downarrow} \rbrace$, and the PW basis, resulting in a combined basis: $\lbrace \ket{\uparrow}, \ket{\downarrow} \rbrace \otimes \lbrace \text{PW basis} \rbrace$.

In this basis, the Bloch wavefunction \( \psi_{n\vec{k}}(\vec{r}) \) can be expanded as:
\begin{equation}
	\psi_{n\vec{k}}(\vec{r}) = \sum_{j,\sigma} C^{\sigma}_{n\vec{k},j} e^{i(\vec{k}+\vec{G}_j) \cdot \vec{r}},
 \end{equation}
with $\bra{\vec{k} + \vec{G}_i} \ket{\vec{k} + \vec{G}_j} = \delta_{ij}$. Here, $\sigma = \uparrow, \downarrow$ represents the spin index. In the case of NSOC (without SOC), this index is ignored, which will be the default assumption in the following discussion. $C^{\sigma}_{n\vec{k},j}$ and $\vec{G}_j$ denote the $j$-th coefficient and the $j$-th reciprocal lattice vector for the $n$-th energy band and spin $\sigma$ at momentum $\vec{k} $, as obtained from VASP calculations.

If we consider a unitary MSG symmetry \( g = \{ \alpha | \vec{t} \} \) of the little group at \( \vec{k} \) acting on the Bloch wavefunction \( \psi_{n\vec{k}}(\vec{r}) \), then the following holds:
\begin{equation}
	\{ \alpha | \vec{t} \} \psi_{n\vec{k}}(\vec{r}) = e^{-i\vec{k} \cdot \vec{t}} \sum_{j, \sigma} C^{\sigma}_{n\vec{k}, j} e^{-i\vec{G}_{j'} \cdot \vec{t}} e^{i (\vec{k} + \vec{G}_{j'}) \cdot \vec{r}},
\end{equation}
where \( \vec{G}_{j'} = \alpha (\vec{k} + \vec{G}_j) - \vec{k} \). The operation matrix acting on the basis can be expressed as \( U(g) = U_{\text{spin}}(g) \otimes U_{\text{space}}(g) \), where \( U_{\text{spin}}(g) \) is the rotation matrix associated with \( g \) acting in spin space, and
\begin{equation}
	U^{j'j}_{\text{space}}(g) = e^{-i\vec{k} \cdot \vec{t}} e^{-i\vec{G}_{j'} \cdot \vec{t}}.
\end{equation}

Consequently, we have:
\begin{equation}
	U(g) C_{n} = \sum_{m} \rho(g)_{mn} C_{m},
\end{equation}
where $\rho(g)$ is the representation matrix of $g$, and $C_{m}$ is the coefficient vector corresponding to the $m$-th energy band. If we define $C = (C_1, \dots, C_N)$ ($N$ is the total number of energy bands), the representation matrix $\rho(g)$ can be computed as:
\begin{equation}
	\rho(g) = (C^{\dagger} C)^{-1} C^{\dagger} U(g) C.
\end{equation}

Since \( [H, g] = 0 \), we have the equation:
	\[
	H g \psi_{n\vec{k}}(\vec{r}) = E_{n\vec{k}} g \psi_{n\vec{k}}(\vec{r}),
	\]
	which implies that the operator \( g \) only transforms the state \( \psi_{n\vec{k}}(\vec{r}) \) into a degenerate state, \( g \psi_{n\vec{k}}(\vec{r}) \). If \( g \) has order \( p \), i.e., \( g^p = 1 \), then applying \( g \) transforms the state \( \psi_{n\vec{k}}(\vec{r}) \) into the degenerate states \( g \psi_{n\vec{k}}(\vec{r}) \), \( g^2 \psi_{n\vec{k}}(\vec{r}) \), and so on up to \( g^{p-1} \psi_{n\vec{k}}(\vec{r}) \). These states are degenerate and related by \( g \), which leads to the representation matrix  \( \rho(g) \) being block-diagonal. To find the character of \( g \) for the energy band \( E_{n\vec{k}} \), we compute the trace of the block-diagonal term corresponding to \( \psi_{n\vec{k}}(\vec{r}) \). Similarly, for any other unitary operation, we can apply the same procedure. In this way, we can determine the characters of the unitary operations in the little group of \( \vec{k} \) for the energy band \( E_{n\vec{k}} \), and consequently obtain the (co)irrep of the energy band \( E_{n\vec{k}} \) based on the (co)irreps of the little group at \( \vec{k} \), as provided by our package \texttt{ToMSGKpoint}.

\section{Installation and Usage}\label{3}
\subsection{Installation}\label{3.1}
To install the \texttt{ToMSGKpoint} package, simply unzip \texttt{ToMSGKpoint.zip} in any folder. Then, open a new Mathematica notebook and run \texttt{Import["Dir\textbackslash ToMSGKpoint.wl"]}, where \texttt{Dir} is the path of the directory containing \texttt{ToMSGKpoint.wl}. \texttt{Import["Dir\textbackslash ToMSGKpoint.wl"]} will first prompt an input box: "Please input the directory of \texttt{ToMSGKpoint.mx}:", where you enter \texttt{"Dir1"}, \texttt{Dir1} is the path of the directory containing \texttt{ToMSGKpoint.mx}. Then, another input box will appear with the prompt: "Please input the working directory of the given crystal structure:", where the working directory is the location where operations for a given crystal structure are performed. To avoid errors, the working directory for different crystal structures should be distinct. Next, another prompt will appear: "Please input a maximal integer for \texttt{lm} (1 might be enough):". Here, \texttt{lm} is a positive integer, and entering \texttt{1} is sufficient in many cases.

\subsection{Usage}\label{3.2}
As mentioned in Sec. \ref{1}, we will sequentially explain how to obtain the MSG operations, how to transform the customized structure convention to the BC convention, how to extract information related to HSPs, HSLs, and HSPLs, and how to use first-principles calculations with VASP to obtain the (co)irreps of every band at HSPs, HSLs, and HSPLs, as well as the distribution of (co)irreps of valence bands across the (co)irreps of little groups at all HSPs, HSLs, and HSPLs.

First, we need to prepare the material's \texttt{POSCAR} file. It is important to note that the crystal structure provided in the POSCAR file must be the smallest unit cell, i.e., the primitive unit cell. By using the \texttt{Getstruct} function outlined below, we can generate a \texttt{struct.mx} file in the specified \texttt{workingdir} (the working directory mentioned earlier). Alternatively, this \texttt{workingdir} can be set to any other directory, allowing you to apply all the operations discussed in this section to a different crystal structure within that folder. The \texttt{struct.mx} file is created from the \texttt{POSCAR} file for use in subsequent operations. The structure of the \texttt{Getstruct} function is as follows:
\begin{tcolorbox}[colback=yellow!10!white, width=0.8\columnwidth]
	\texttt{Getstruct[poscar, workingdir]}
\end{tcolorbox}
In this setup, replace \texttt{poscar} with the path to the material's \texttt{POSCAR} file and \texttt{workingdir} with the path to the working directory.

The structure of the generated \texttt{struct.mx} is as follows:
\[
\left\{ \{\vec{a}_1, \vec{a}_2, \vec{a}_3\}, \left\{ \{ \text{element}, \text{position}, \text{magnetic moment} \} \dots \right\} \right\}
\]
where \(\vec{a}_1, \vec{a}_2, \vec{a}_3\) represent the primitive lattice vectors, \textit{element} refers to the type of element, \textit{position} is the fractional coordinate of the element relative to \(\vec{a}_1, \vec{a}_2, \vec{a}_3\), and \textit{magnetic moment} is the magnetic moment of the element in the Cartesian coordinate system. The default magnetic moment is \(\{0, 0, 0\}\) for every element. If you wish to study magnetic materials, you need to manually specify the magnetic moment for each element in \texttt{struct.mx}. The explanation of the struct.mx example can refer to Figures \ref{s1}, \ref{s2}, and \ref{s3} in Section \ref{5} for guidance. In the following, we use \texttt{struct} to represent the full data of the file \texttt{struct.mx}.

To obtain the MSG operations for the given structure in \texttt{struct.mx}, we use the \texttt{GetMSGOP} function. The structure of the function is as follows:
\begin{tcolorbox}[colback=yellow!10!white, width=0.8\columnwidth]
	\texttt{GetMSGOP[$a1a2a3$, atoms, $\epsilon$, lm1]}
\end{tcolorbox}
where $a1a2a3=\{\vec{a}_1, \vec{a}_2, \vec{a}_3\}$, $\texttt{atoms}=\left\{ \{ \text{element}, \text{position}, \text{magnetic moment} \} \dots \right\}$, \(\epsilon\) is the precision for finding MSG operations, and \texttt{lm1} is a positive integer, typically set to 1. It is worth noting that when you select a different \(\epsilon\), such as \(0.01 \rightarrow 0.05\), you might have more MSG operations. This is because the symmetry obtained here can be considered an approximate symmetry, which changes as you adjust the precision. You can adjust this \(\epsilon\) according to your needs.

The output of the \texttt{GetMSGOP} function is in the following format: \{\text{ty}, \{\text{op1}, \text{op2}\}\}, where \text{ty} represents the type of the MSG for the given structure, \text{op1} are the unitary MSG operations for the given structure, and \text{op2} are the unitary part \( U \) of the anti-unitary MSG operations. Specifically, the anti-unitary MSG operations can be written as \( UT \), where \( T \) denotes time-reversal symmetry.

Then, to obtain the transformation from the customized primitive crystal structure convention to the BC convention, we use the \texttt{GetMSG} function:
\begin{tcolorbox}[colback=yellow!10!white, width=0.8\columnwidth]
	\texttt{GetMSG[a1a2a3, $\{\text{ty},\{\text{op1},\text{op2}\}\}$, $\epsilon$, lm2]}
\end{tcolorbox}
where \texttt{lm2} is a positive integer, typically set to 1. If an error occurs when running the \texttt{GetMSGOP} function, you can gradually increase the value of \texttt{lm2} and then rerun the function.

The output of the \texttt{GetMSG} function is in the following format: $\{\text{flag}, \vec{\delta},\text{sg}, P, Ur, U_{xyz}, \text{ty}, \{\text{ty}, \text{X.Y}\}\}$, where $\vec{\delta}, P$ and $U_{xyz}$ are the parameters required for the transformation from the customized structure convention to the BC convention, as described in Sec. \ref{2}. The symbol X.Y represents the MSG name for this structure in the BC convention.

Third, to obtain the little groups and (co)irreps of the little groups of HSPs, HSLs and HSPLs, we use the \texttt{ToHSPInfo}, \texttt{ToHSLInfo} and \texttt{ToHSPLInfo} functions:
\begin{tcolorbox}[colback=yellow!10!white, width=0.8\columnwidth]
	\texttt{ToHSPInfo[sg, $\vec{\delta}$, P, Uxyz, so]}\\
	\texttt{ToHSLInfo[a1a2a3, sg, $\vec{\delta}$, P, Uxyz, so]}\\
	\texttt{ToHSPLInfo[a1a2a3, sg, $\vec{\delta}$, P, Uxyz, so]}
\end{tcolorbox}
where \texttt{sg, $\vec{\delta}$, P, Uxyz}  correspond to the third, second, fourth, and sixth positions, respectively, in the output of the \texttt{GetMSG} function, and $\texttt{so}=0(1)$ indicates NSOC (SOC).

For convenience and to facilitate subsequent first-principles calculations, we also define the \texttt{Getgkinfo}, \texttt{ShowMSGop}, \texttt{ShowTrantoBC}, and \texttt{ShowKpointinfo} functions. The structure of the \texttt{Getgkinfo} function is as follows:
\begin{tcolorbox}[colback=yellow!10!white, width=0.8\columnwidth]
	\texttt{Getgkinfo[workingdir, struct, $\epsilon$, lm1, lm2, py, nl, pa, so]}
\end{tcolorbox}
Where \texttt{py} represents a choice between HSP, HSL, and HSPL: \texttt{py = 1} for HSPs, \texttt{py = 2} for HSLs, and \texttt{py = 3} for HSPLs. The parameter \texttt{nl} specifies which HSL or HSPL is selected when considering HSLs or HSPLs, while \texttt{pa} indicates how many points to sample along the selected HSL or a $\vec{k}$-point path of HSPL. When considering HSPs, the parameters \texttt{nl} and \texttt{pa} do not have any effect, so you can set them to 1 when \texttt{py = 1}. \texttt{lm1} and \texttt{lm2} correspond to \texttt{lm1} in the \texttt{GetMSGop} function and \texttt{lm2} in the \texttt{GetMSG} function, respectively.

The \texttt{Getgkinfo} function generates the following files: \texttt{msgop.mx}, \texttt{transformation.mx}, \texttt{gkHSPinfo.mx}, \texttt{gkHSLinfo.mx} (\texttt{gkHSLinfo$\lbrace$nl$\rbrace$.mx} for the \texttt{nl}-th HSL), \texttt{gkHSPLinfo.mx} (\texttt{gkHSPLinfo$\lbrace$nl$\rbrace$.mx} for the \texttt{nl}-th HSPL), and the KPOINTS file for VASP non-self-consistent first-principles electronic wavefunction calculations. Among these:
\begin{itemize}
	\item \texttt{msgop.mx} contains the output of the \texttt{GetMSGOP} function (the \texttt{ShowMSGop} function visualizes the corresponding MSG operations),
	\item \texttt{transformation.mx} contains the output of the \texttt{GetMSG} function (the \texttt{ShowTrantoBC} function visualizes the corresponding transformation),
	\item \texttt{gkHSPinfo.mx}, \texttt{gkHSLinfo.mx}, or \texttt{gkHSPLinfo.mx} contains the output of the \texttt{ToHSPInfo}, \texttt{ToHSLInfo}, or \texttt{ToHSPLInfo} function (the \texttt{ShowKpointinfo} function visualizes the HSP, HSL, and HSPL information).
\end{itemize}

Finally, to obtain the (co)irrep information of the band structure at HSPs, HSLs, and HSPLs, as well as the distribution of (co)irreps of all valence electrons across the (co)irreps of the little groups at these points, we use the \texttt{GetKpointTop} function. Before using it, perform the first-principles calculations and place the WAVECAR and OUTCAR files from the final non-self-consistent band structure calculation into the \texttt{workingdir}. The structure of the \texttt{GetKpointTop} function is as follows:

\begin{tcolorbox}[colback=yellow!10!white, width=0.8\columnwidth]
	\texttt{GetKpointTop[workingdir, $\epsilon_1$, $\epsilon_2$, py, nl, top, so]}
\end{tcolorbox}
Where:
\begin{itemize}
	\item $\epsilon_1$ controls the precision for determining whether two bands are degenerate,
	\item $\epsilon_2$ controls the precision for determining whether the occurrence of (co)irreps is close to an integer,
	\item \texttt{top = 0} specifies that the function computes the (co)irrep information of the band structure at HSPs, HSLs, or HSPLs. The \texttt{GetKpointTop} function generates two files: \texttt{kpoint.mx} and \texttt{HSPbandirrep.mx} for HSP, while \texttt{HSLbandirrep$\lbrace$\texttt{nl}$\rbrace$.mx} is for HSL, and \texttt{HSPLbandirrep$\lbrace$\texttt{nl}$\rbrace$.mx} is for HSPL.
	\item \texttt{top = 1} specifies the distribution of (co)irreps of all valence bands across the (co)irreps of little groups of HSPs. The \texttt{GetKpointTop} function generates a file, \texttt{HSPTop.mx}.
	This file can be used as an input file for the \texttt{CalTopoEvol} package \cite{tang} to calculate the topological properties of the material.
\end{itemize}

It is important to note that the files \texttt{HSPbandirrep.mx}, \texttt{HSLbandirrep$\lbrace$\texttt{nl}$\rbrace$.mx}, and \texttt{HSPLbandirrep$\lbrace$\texttt{nl}$\rbrace$.mx} contain the (co)irreps for all bands at the HSPs, along the $\vec{k}$-point path on the \texttt{nl}-th HSL, or along the $\vec{k}$-point path at the \texttt{nl}-th HSPL. For certain higher energy levels above the Fermi energy at these HSPs, HSLs, and HSPLs, the little group of these bands may be incorrectly determined, preventing us from obtaining the correct (co)irreps for these bands. As a result, we do not specify the (co)irreps for these bands in the files \texttt{HSPbandirrep.mx}, \texttt{HSLbandirrep$\lbrace$\texttt{nl}$\rbrace$.mx}, or \texttt{HSPLbandirrep$\lbrace$\texttt{nl}$\rbrace$.mx}.

We also define the \texttt{ShowBandirreppoint} and \texttt{ShowBandirrepline} functions to visualize the (co)irreps of energy bands at HSPs, HSLs, or HSPLs. The structures of the \texttt{ShowBandirreppoint} and \texttt{ShowBandirrepline} functions are as follows:
\begin{tcolorbox}[colback=yellow!10!white, width=0.8\columnwidth]
	\texttt{ShowBandirreppoint[workingdir, py, nl, kpoint, band1, band2]}
	\texttt{ShowBandirrepline[workingdir, py, nl, sband, kpoint1, kpoint2]}
\end{tcolorbox}
Here, \texttt{kpoint} denotes the index of the $\vec{k}$-point in the \texttt{kpoint.mx} file, \texttt{band1} specifies the lower bound of the band range to be displayed, \texttt{band2} specifies the upper bound, and \texttt{sband} represents a specific band. The \texttt{ShowBandirreppoint} function displays the (co)irreps of energy bands at the \texttt{kpoint}-th $\vec{k}$-point at the HSPs, HSLs, or HSPLs. The \texttt{ShowBandirrepline} function shows the (co)irrep of the \texttt{sband}-th energy band along the $\vec{k}$-point path at the \texttt{nl}-th HSL or HSPL, from the \texttt{kpoint1}-th to the \texttt{kpoint2}-th $\vec{k}$-point. The path follows the order specified in the \texttt{kpoint.mx} file.

\begin{figure}[H]
	\centering
	\begin{tikzpicture}
		\node[draw=green!50,fill=green!20,very thick,minimum width=2cm,minimum height=1cm,rectangle,align=center ] (1) at (-2,0) {\texttt{Getstruct}};
		\node[draw=green!50,fill=green!20,very thick,minimum width=2cm,minimum height=1cm,rectangle,align=center ] (2) at (2.9,0) {\texttt{Getgkinfo}};
		\draw[->] (1)--(2);
		\node[draw=green!50,fill=green!20,very thick,minimum width=4cm,minimum height=1cm,rectangle,align=center ] (5) at (2.9,-2.2) {\texttt{ShowMSGop},\texttt{ShowTrantoBC},\texttt{ShowKpointinfo}};
		\draw[->] (2)--(5);
		\node at(3.7,-0.9) {\texttt{msgop.mx}};
		\node at(4.5,-1.3) {\texttt{transformation.mx}};
		\node at(1.7,-0.7) {\texttt{gkHSPinfo.mx}};
		\node at(1.7,-1.1) {\texttt{gkHSLinfo.mx}};
		\node at(1.6,-1.5) {\texttt{gkHSPLinfo.mx}};
		\node at(0.5,0.3) {\texttt{struct.mx}};
		\node[draw=green!50,fill=green!20,very thick,minimum width=2cm,minimum height=1cm,rectangle,align=center ] (3) at (7.3,0) {VASP calculation};
		\draw[->] (2)--(3);
		\node at(4.8,0.3) {KPOINTS};
		\node[draw=green!50,fill=green!20,very thick,minimum width=2cm,minimum height=1cm,rectangle,align=center ] (4) at (12,0) {\texttt{GetKpointTop}};
		\draw[->] (3)--(4);
		\node at(9.7,-0.3){OUTCAR};
		\node at(9.7,0.3) {WAVECAR};
		\node[draw=green!50,fill=green!20,very thick,minimum width=2cm,minimum height=1cm,rectangle,align=center ] (6) at (-2,-2.2) {\texttt{workingdir}};
		\draw[->] (6)--(1);
		\node at(-1.2,-1){POSCAR};
		\node[draw=green!50,fill=green!20,very thick,minimum width=1cm,minimum height=1cm,rectangle,align=center ] (7) at (12,-2.2) {\texttt{ShowBandirreppoint}\\ \texttt{ShowBandirrepline}};
		\draw[->] (4)--(7);
		\node at(10.45,-0.7){\texttt{HSPbandirrep.mx}};
		\node at(10.1,-1.1){\texttt{HSLbandirrep$\lbrace$nl$\rbrace$.mx}};
		\node at(10.,-1.5){\texttt{HSPLbandirrep$\lbrace$nl$\rbrace$.mx}};
	\end{tikzpicture}
	\caption{The flowchart represents the entire computational process of the \texttt{ToMSGKpoint} package.}
	\label{fig:1}
\end{figure}
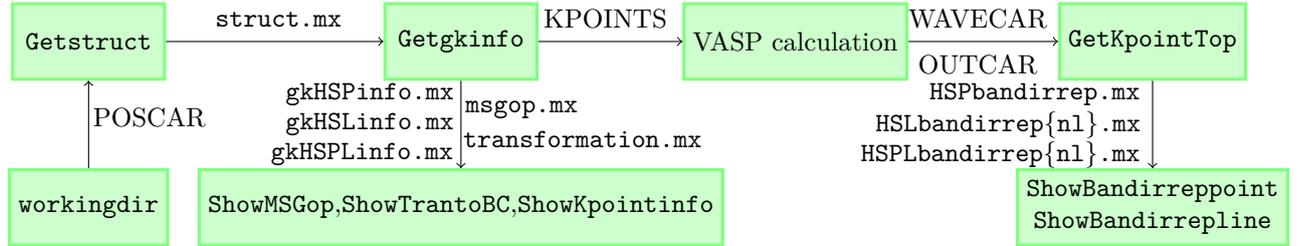

Although the above introduction is quite detailed, only a few simple steps are needed when using it. The entire computational process of the \texttt{ToMSGKpoint} package can be seen in Figure \ref{fig:1}.

\section{Examples}\label{4}
In this section, we demonstrate the use of the \texttt{ToMSGKpoint} package with three examples: the nonmagnetic topological insulator Bi$_2$Se$_3$ (focusing on HSP), Dirac semimetal Na$_3$Bi (focusing on HSL, along the \(k_z\)-direction in Cartesian coordinates), and the AFM topological insulator MnBi$_2$Te$_4$ (focusing on HSP). In all \delete{three} cases, SOC is considered.

\subsection{nonmagnetic topological insulator Bi$_2$Se$_3$}
Bi\(_2\)Se\(_3\) is a typical three-dimensional strong topological insulator, with a Dirac cone present on its surface \cite{b1,b2,b3}. Using this material as an example, we will demonstrate that even when the given structure significantly deviates from the common conventions, we can still obtain all the results that we are interested in.

\begin{figure}[H]
	\centering
	\subfloat[\label{subfig:3a}]{
		\includegraphics[width=0.35\columnwidth]{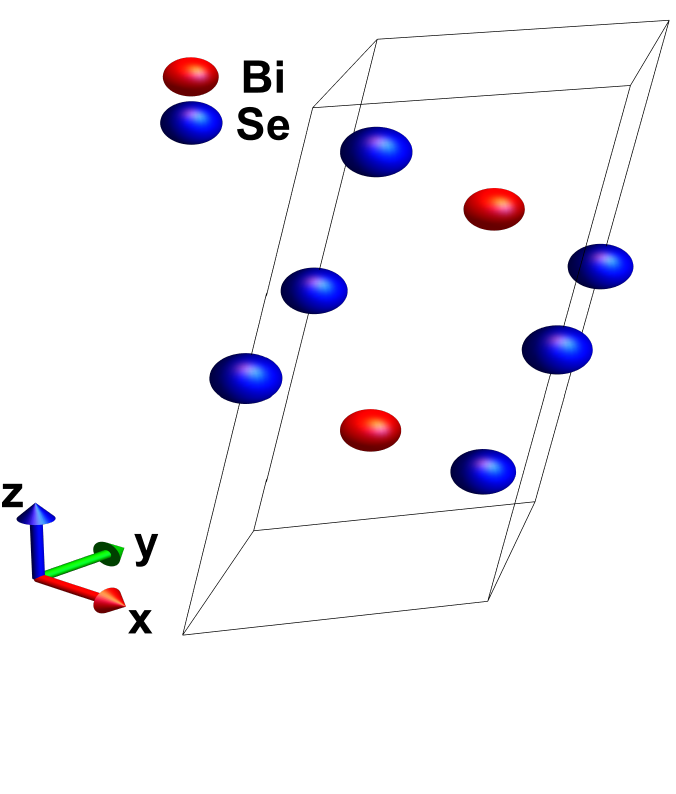}}
	\subfloat[\label{subfig:3b}]{
		\includegraphics[width=0.65\columnwidth]{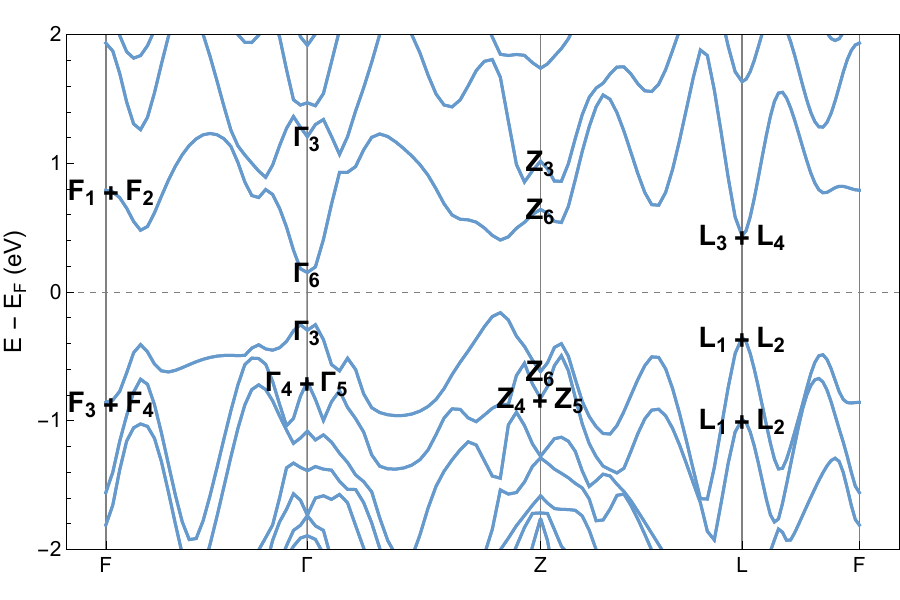}}
	\caption{(a) and (b) display the crystal structure and the energy bands along a closed path in the BZ of Bi$_2$Se$_3$, respectively. Panel (b) also shows the corresponding (co)irreps of the energy bands near the Fermi energy at all HSPs. The coordinates of the HSPs are as follows: $\Gamma : (0,0,0)$, $\text{Z} : (1,1,-0.5)$, $\text{L} : (0,0.5,0)$, and $\text{F} : (0.5,1,-0.5)$.}
	\label{fig:3}
\end{figure}

First, we create a new directory, \texttt{workingdir}, and then use the \texttt{Getstruct} function to generate the \texttt{struct.mx} file. For the non-magnetic material Bi$_2$Se$_3$, we do not need to modify the \texttt{struct.mx} file. The content of the \texttt{struct.mx} file is shown in Figure \ref{s1}:

\begin{figure}[H]
	\centering
	\includegraphics[width=0.9\columnwidth]{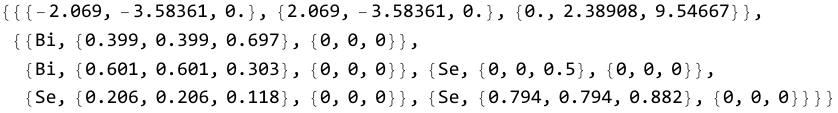}
	\caption{The structure data of Bi$_2$Se$_3$ in the \texttt{struct.mx} file.}
	\label{s1}
\end{figure}
From Figure \ref{s1}, we know that the primitive lattice vectors of the given Bi$_2$Se$_3$ structure are $\vec{a}_1=(-2.069, -3.584, 0)$, $\vec{a}_2=(2.069, -3.584, 0)$, $\vec{a}_3=(0, 2.389, 9.547)$ in units of \AA\ in the Cartesian coordinate system $Oxyz$. Bi is at $(0.399, 0.399, 0.697)$ and $(0.601, 0.601, 0.303)$; Se is at $(0, 0, 0.5)$, $(0.206, 0.206, 0.118)$, and $(0.794, 0.794, 0.882)$ relative to $\vec{a}_1$, $\vec{a}_2$, and $\vec{a}_3$. The crystal structure can be seen in Figure \ref{subfig:3a}. The given $\lbrace$$\vec{a}_1, \vec{a}_2, \vec{a}_3$$\rbrace$ are far from the standard primitive lattice vectors $\lbrace$(-2.095, -1.209, 10.137), (2.095, -1.209, 10.137), (0, 2.419, 10.137)$\rbrace$, and as shown below, we only need to set \texttt{lm2 = 4} when applying \texttt{Getgkinfo} function to obtain the corresponding MSG operations, transformation to BC convention, and the unitary operations of the little group at HSPs, along with its (co)irreps.

\begin{figure}[htbp]
	\centering
	\includegraphics[width=0.9\columnwidth]{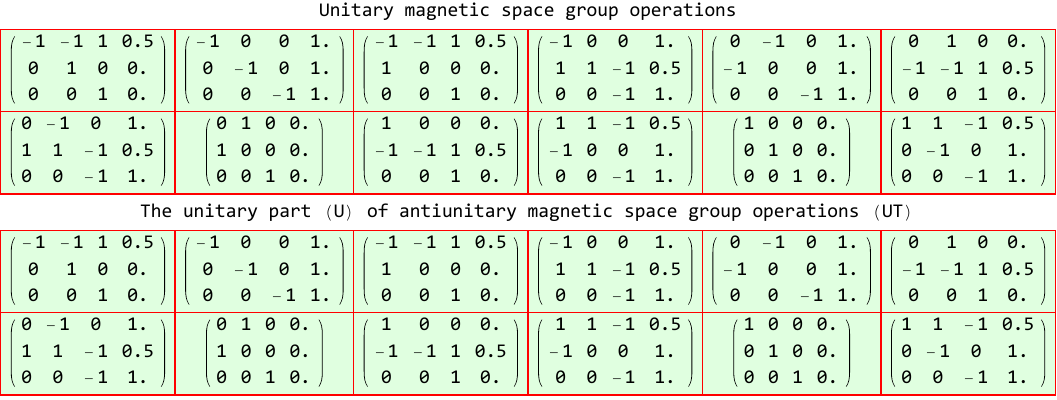}
	\caption{This figure shows the output of \texttt{ShowMSGop[workingdir]}, which lists all the MSG operations for the given Bi$_2$Se$_3$ structure.
	}
	\label{b1}
\end{figure}

We use the \texttt{Getgkinfo[workingdir, struct, 0.001, 1, 4, 1, 1, 1, 1]} in \texttt{workingdir} to generate four files: \texttt{msgop.mx}, \texttt{transformation.mx}, \texttt{gkHSPinfo.mx}, and the KPOINTS file. The \texttt{ShowMSGop} function is then used to visualize the corresponding MSG operations, see Figure \ref{b1}.
We use the first unitary MSG operation
\[
\begin{pmatrix}
	-1 & -1 & 1 & 0.5 \\
	0 & 1 & 0 & 0 \\
	0 & 0 & 1 & 0 \\
\end{pmatrix}
\]
in Figure \ref{b1} to illustrate the meaning of the matrix in the figure. The first three columns represent the rotation operation, and the fourth column represents the translation operation. In other words, if we represent this operation as \(\{ \alpha | \vec{t} \}\), then
\[
\alpha = \begin{pmatrix}
	-1 & -1 & 1 \\
	0 & 1 & 0 \\
	0 & 0 & 1 \\
\end{pmatrix}, \quad \vec{t} = (0.5, 0, 0).
\]
This matrix will leave the Bi atom at \((0.399, 0.399, 0.697)\) unchanged, move the Bi atom at \((0.601, 0.601, 0.303)\) to the position \((-0.399, 0.601, 0.303)\), i.e., \((0.601, 0.601, 0.303) - \vec{a}_1\), move the Se atom at \((0, 0, 0.5)\) to the position \((1, 0, 0.5)\), keep the Se atom at \((0.206, 0.206, 0.118)\) unchanged, and move the Se atom at \((-0.206, 0.794, 0.882)\) to the position \((0.794, 0.794, 0.882)\). The other MSG operations are similar.

\begin{figure}[htbp]
	\centering
	\includegraphics[width=0.9\columnwidth]{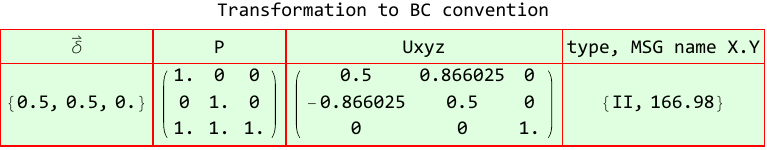}
	\caption{This figure shows the output of \texttt{ShowTrantoBC[workingdir]}, which provides the three parameters $\vec{\delta}$, $P$, and $U_{\text{xyz}}$ mentioned earlier, describing the transformation from the given crystal structure convention of Bi$_2$Se$_3$ to the BC convention.}
	\label{b2}
\end{figure}

After discussing the MSG operations, we will describe the transformation from the customized crystal structure convention to the BC convention and we use the \texttt{ShowTrantoBC} function to visualize this transformation, as shown in Figure \ref{b2}.

\begin{figure}[htbp]
	\centering
	\includegraphics[width=0.9\columnwidth]{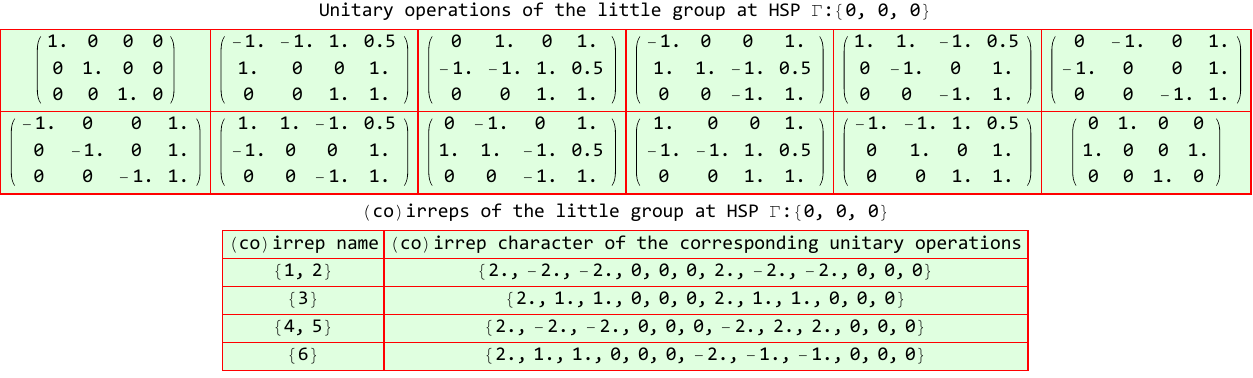}
	\caption{This figure displays the output of \texttt{ShowKpointinfo[workingdir, 1, 1]}. For each unitary MSG operation $g$, we show only the first character of the double-valued representation. The complete results can be found in the \texttt{gkHSPinfo.mx} file in the working directory.}
	\label{b3}
\end{figure}

As mentioned earlier, after obtaining this transformation, we can indirectly determine the little group of HSPs and their (co)irreps. Using the function \texttt{ShowKpointinfo[workingdir, 1, 1]} (where the first 1 refers to the HSP and the second 1 refers to the first HSP in the \texttt{gkHSPinfo.mx} file in the working directory), we demonstrate the unitary MSG operations of the little group at the HSP $\Gamma: (0,0,0)$ and the character table of the unitary MSG operations of its (co)irreps, as shown in Figure \ref{b3}.

Then, we performed VASP calculations. After completing the self-consistent calculations, we proceeded with non-self-consistent band structure calculations, using the KPOINTS file generated earlier by the \texttt{Getgkinfo} function. Before this step, it is necessary to delete the WAVECAR file obtained from the self-consistent calculations to prepare for reading the wavefunctions for the non-self-consistent band structure calculations. Once the calculations are complete, place the WAVECAR and OUTCAR files from the non-self-consistent process into the working directory. Subsequently, using the command \texttt{GetKpointTop[workingdir, 0.00001, 0.01, 1, 1, 0, 1]}, we computed and generated the \texttt{HSPbandirrep.mx} file. Figure \ref{subfig:3b} displays the energy bands along a closed path in the Brillouin zone (BZ), as well as the (co)irreps of energy bands near the Fermi energy at all HSPs of Bi$_2$Se$_3$.

\begin{figure}[H]
	\centering
	\includegraphics[width=0.75\columnwidth]{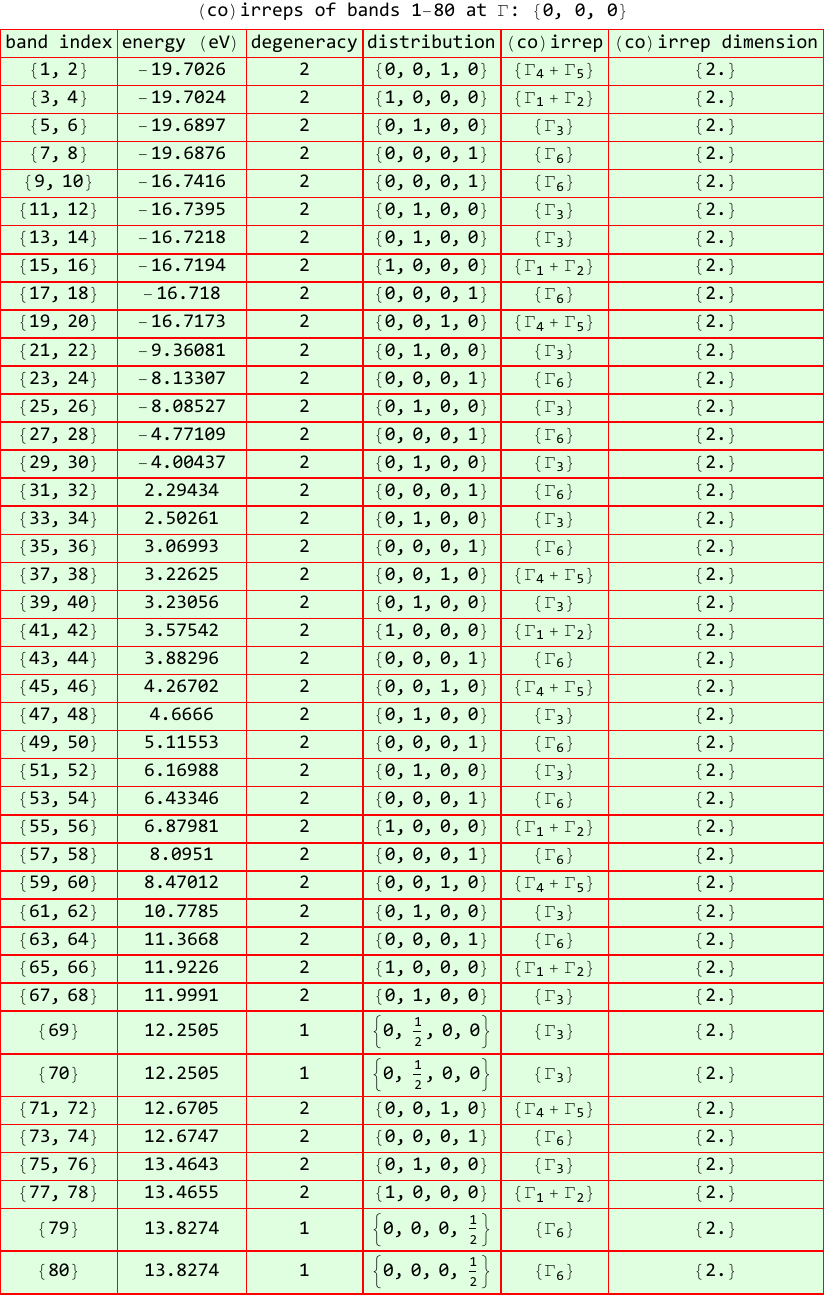}
	\caption{This figure shows the results of \texttt{ShowBandirreppoint[workingdir, 1, 1, 1, 1, 80]}. In the figure, $\Gamma_1 + \Gamma_2, \Gamma_4 + \Gamma_5$ represent the (co)irreps of the energy bands, which correspond to the coirreps $\lbrace 1, 2 \rbrace$, $\lbrace 4, 5 \rbrace$ at the HSP $\Gamma : (0,0,0)$ in Figure \ref{b3}. Similarly, $\Gamma_3, \Gamma_6$ represent the (co)irreps of the energy bands, corresponding to the irreps $\lbrace 3 \rbrace$, $\lbrace 6 \rbrace$ at the HSP $\Gamma : (0,0,0)$ in Figure \ref{b3}.}
	\label{fig:33}
\end{figure}

To demonstrate that \texttt{ToMSGKpoint} can compute the (co)irreps of all energy bands at HSPs, along HSLs, and on HSPLs, we use Bi$_2$Se$_3$ as an example. In Figure \ref{fig:33}, the command \texttt{ShowBandirreppoint[workingdir, 1, 1, 1, 1, 80]} is used to display the (co)irreps of a total of 80 energy bands at the HSP $\Gamma : (0,0,0)$.

Finally, we use the function \texttt{GetKpointTop[workingdir, 0.00001, 0.01, 1, 1, 1, 1]} to generate the file \texttt{HSPTop.mx}, which serves as the input file for the package \texttt{CalTopoEvol}. By using \texttt{CalTopoEvol}, we find that the symmetry indicator group for Bi$_2$Se$_3$, considering SOC, is $\mathbb{Z}_2 \times \mathbb{Z}_4$, with a symmetry indicator of $(0,1)$. This result indicates that Bi$_2$Se$_3$ is a strong topological insulator.

\subsection{AFM topological insulator MnBi$_2$Te$_4$}
The AFM material MnBi$_2$Te$_4$, recently predicted as the first intrinsic AFM topological insulator, has attracted significant attention both experimentally and theoretically \cite{m1,m2,m3,m4,m5}. Here, we take it as an example to demonstrate that the package \texttt{ToMSGKpoint} is also applicable to magnetic materials.

\begin{figure}[H]
	\centering
	\subfloat[\label{subfig:4a}]{
		\includegraphics[width=0.35\columnwidth]{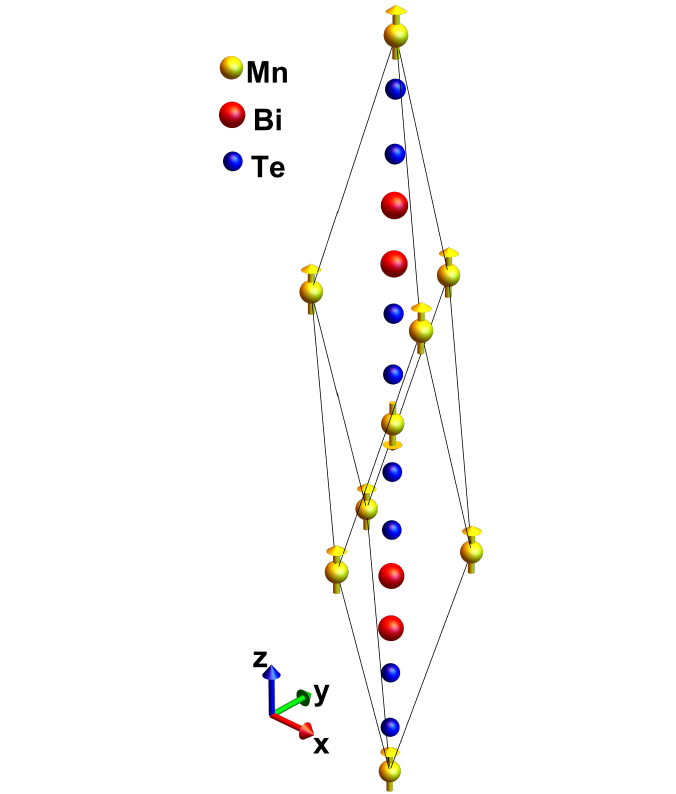}}
	\subfloat[\label{subfig:4b}]{
		\includegraphics[width=0.65\columnwidth]{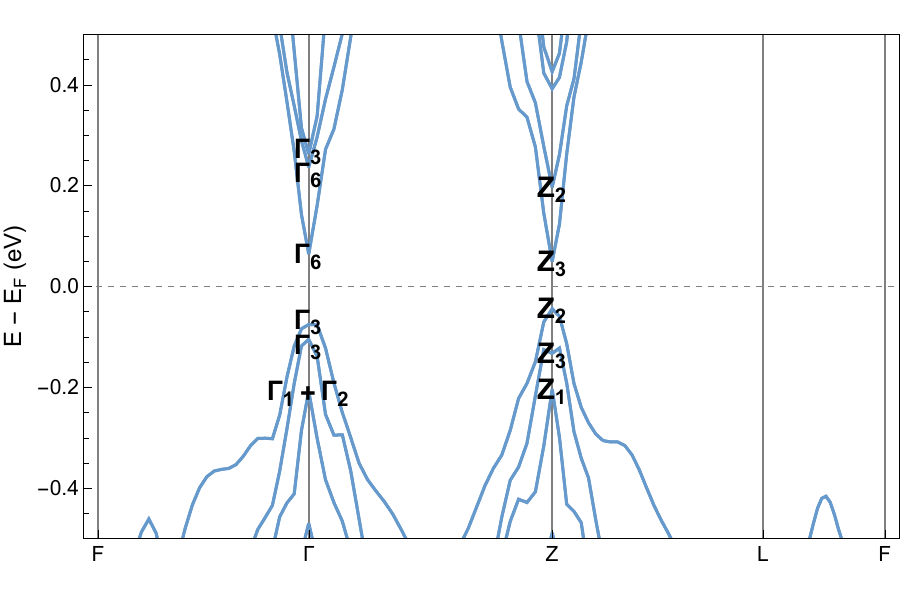}}
	\caption{(a) and (b) display the crystal structure and the energy bands ($U = 3 \,\mathrm{eV}$) along a closed path in the BZ of MnBi$_2$Te$_4$, respectively. In panel (b), we also present the corresponding (co)irreps of the energy bands near the Fermi energy at all HSPs. The coordinates of the HSPs are as follows: $\Gamma :(0,0,0)$, $\text{Z} : (0.5,0.5,-0.5)$, $\text{L} : (0,0.5,0)$, and $\text{F}:(0,0.5,-0.5)$.}
	\label{fig:4}
\end{figure}

First, we create a new directory, \texttt{workingdir}, and then use the \texttt{Getstruct} function to generate the \texttt{struct.mx} file. For the AFM material MnBi$_2$Te$_4$, we need to modify the \texttt{struct.mx} file. The content of the modified \texttt{struct.mx} file is shown in Figure \ref{s2}:

\begin{figure}[H]
	\centering
	\includegraphics[width=1\columnwidth]{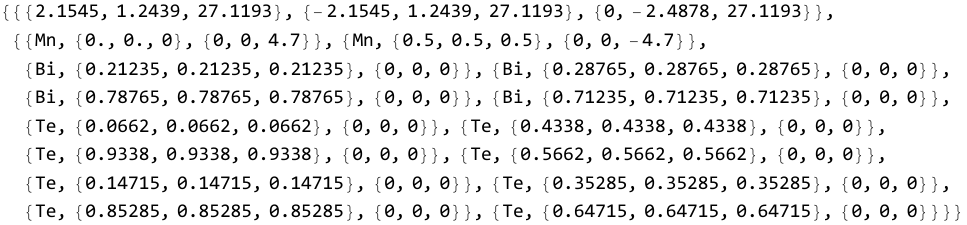}
	\caption{The structure data of MnBi$_2$Te$_4$ in the modified \texttt{struct.mx} file.}
	\label{s2}
\end{figure}
From Figure \ref{s2}, we know that the primitive lattice vectors are $\vec{a}_1=(2.155, 1.244, 27.119)$, $\vec{a}_2=(-2.155, $\\$1.244, 27.119)$, $\vec{a}_3=(0, -2.488, 27.119)$ in units of \AA\ in the Cartesian coordinate system $Oxyz$. Mn is at $(0, 0, 0)$ and $(0.5, 0.5, 0.5)$; Bi is at $(0.212, 0.212, 0.212)$, $(0.288, 0.288, 0.288)$, $(0.788, 0.788, 0.788)$, and $(0.712, 0.712, 0.712)$; Te is at $(0.066, 0.066, 0.066)$, $(0.434, 0.434, 0.434)$, $(0.934, 0.934, 0.934)$, $(0.566, 0.566, 0.566)$, $(0.147, 0.147, 0.147)$, $(0.353, 0.353, 0.353)$, $(0.853, 0.853, 0.853)$, and $(0.647, 0.647, 0.647)$ relative to $\vec{a}_1$, $\vec{a}_2$, and $\vec{a}_3$. The magnetic moment (in units of $\mu_B$) of the magnetic atom Mn is $(0, 0, 4.7)$ at $(0, 0, 0)$ and $(0, 0, -4.7)$ at $(0.5, 0.5, 0.5)$ in the Cartesian coordinate system $Oxyz$. The crystal structure and magnetic structure can be seen in Figure \ref{subfig:4a}.

\begin{figure}[H]
	\centering
	\includegraphics[width=1\columnwidth]{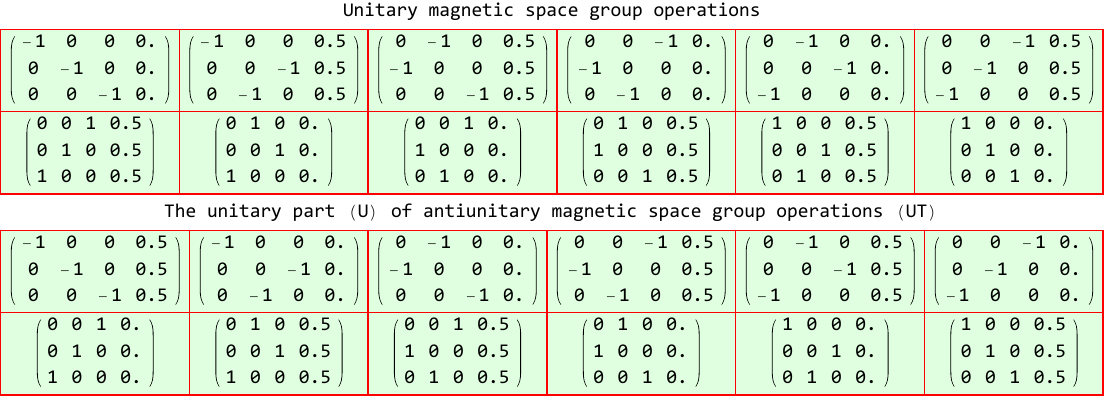}
	\caption{This figure shows the output of \texttt{ShowMSGop[workingdir]}, which lists all the MSG operations for the given MnBi$_2$Te$_4$ structure.}
	\label{m1}
\end{figure}

We use the \texttt{Getgkinfo[workingdir, struct, 0.001, 1, 1, 1, 1, 1, 1]} in \texttt{workingdir} to generate four files: \texttt{msgop.mx}, \texttt{transformation.mx}, \texttt{gkHSPinfo.mx}, and the KPOINTS file. The \texttt{ShowMSGop} function is then used to visualize the corresponding MSG operations, see Figure \ref{m1}.
We use the first anti-unitary MSG operation in Figure \ref{m1} as an example to demonstrate how the magnetic structure is preserved under the MSG operations listed in the same figure. Representing this operation as \(\{ \alpha | \vec{t} \}T\), we obtain:
\[
\alpha = \begin{pmatrix}
	-1 & 0 & 0 \\
	0 & -1 & 0 \\
	0 & 0 & -1
\end{pmatrix}, \quad \vec{t} = (0.5, 0.5, 0.5).
\]
This operation maps the Mn atom at \((0, 0, 0)\) to the Mn atom at \((0.5, 0.5, 0.5)\). Furthermore, due to the action of time-reversal symmetry \(T\), the magnetic moment is reversed in sign.

\begin{figure}[H]
	\centering
	\includegraphics[width=1\columnwidth]{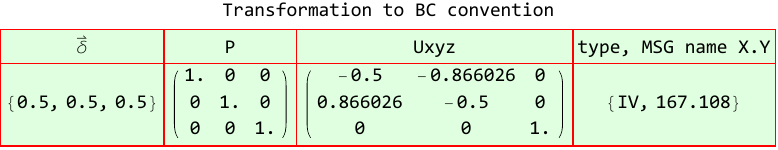}
	\caption{This figure shows the output of \texttt{ShowTrantoBC[workingdir]}, which provides the three parameters $\vec{\delta}$, $P$, and $U_{\text{xyz}}$ mentioned earlier, describing the transformation from the given crystal structure convention of MnBi$_2$Te$_4$ to the BC convention.}
	\label{m2}
\end{figure}

After we talk about the MSG operations, let us give the transformation from the customized crystal structure convention to the BC convention and use the \texttt{ShowTrantoBC} function to visualize this transformation, as shown in Figure \ref{m2}.

\begin{figure}[htbp]
	\centering
	\includegraphics[width=0.9\columnwidth]{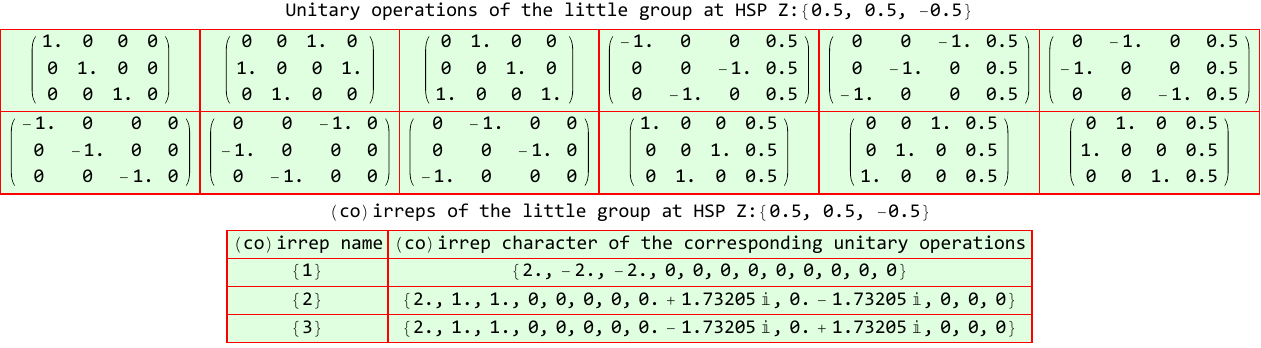}
	\caption{This figure displays the output of \texttt{ShowKpointinfo[workingdir, 1, 2]}. For each unitary MSG operation $g$, we show only the first character of the double-valued representation. The complete results can be found in the \texttt{gkHSPinfo.mx} file in the working directory.}
	\label{m3}
\end{figure}

Using the function \texttt{ShowKpointinfo[workingdir, 1, 2]} (where the first argument, 1, refers to the HSP, and the second argument, 2, refers to the second HSP in the \texttt{gkHSPinfo.mx} file within the working directory), we demonstrate the unitary MSG operations of the little group at the HSP $\text{Z}: (0.5, 0.5, -0.5)$, along with the character table of its (co)irreps, as shown in Figure \ref{m3}.

Then, we proceeded with first-principles calculations. For MnBi$_2$Te$_4$, we performed DFT+U calculations, where we set $U = 3 \,\mathrm{eV}$ in this study. After the DFT+U calculations, we used the command \texttt{GetKpointTop[workingdir, 0.00001, 0.01, 1, 1, 0, 1]} to compute and generate the \texttt{HSPbandirrep.mx} file. Figure \ref{subfig:4b} displays the energy bands along a closed path in the BZ as well as the (co)irreps of energy bands near the Fermi energy at all HSPs of MnBi$_2$Te$_4$.

\begin{figure}[H]
	\centering
	\includegraphics[width=0.9\columnwidth]{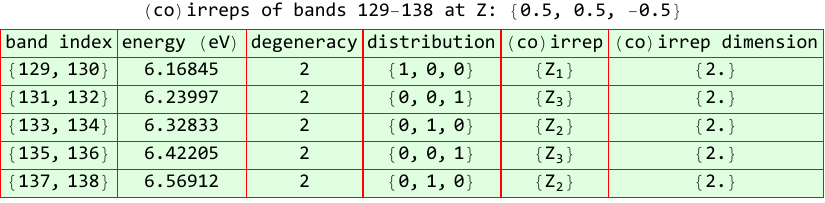}
	\caption{This figure shows the results of \texttt{ShowBandirreppoint[workingdir, 1, 1, 1, 129, 138]}. In the figure, $\text{Z}_1, \text{Z}_2, \text{Z}_3$ represent the (co)irreps of the energy bands, which correspond to the irreps $\lbrace 1 \rbrace$, $\lbrace 2 \rbrace$ and $\lbrace 3 \rbrace$ at the HSP $\text{Z} : (0.5,0.5,-0.5)$ in Figure \ref{m3}.}
	\label{fig:43}
\end{figure}

In Figure \ref{fig:43}, the function \texttt{ShowBandirreppoint[workingdir, 1, 1, 1, 129, 138]} is used to display the (co)irreps of five two-degenerate energy bands at the HSP $\text{Z} :(0.5,0.5,-0.5)$, as shown in Figure \ref{subfig:4b}.

Finally, we use the function \texttt{GetKpointTop[workingdir, 0.00001, 0.01, 1, 1, 1, 1]} to generate the file \texttt{HSPTop.mx}, which serves as the input file for the package \texttt{CalTopoEvol}. By using \texttt{CalTopoEvol}, we find that the symmetry indicator group for AFM material MnBi$_2$Te$_4$, considering SOC, is $\mathbb{Z}_2$, with a symmetry indicator of $1$. This result indicates that AFM material MnBi$_2$Te$_4$ is a magnetic topological insulator.

\subsection{nonmagnetic Dirac semimetal Na$_3$Bi}
Na$_3$Bi is a three-dimensional topological Dirac semimetal, characterized by a pair of Dirac points located along the \( k_z \)-axis in the bulk states \cite{na1,na2}. Here, we use it as an example to demonstrate that the package \texttt{ToMSGKpoint} can compute the (co)irreps of energy bands along HSLs. By analyzing the inversion of the (co)irreps of two double-degenerate bands, the presence of the four-degenerate Dirac nodes can be predicted.

\begin{figure}[H]
	\centering
	\subfloat[\label{subfig:5a}]{
		\includegraphics[width=0.35\columnwidth]{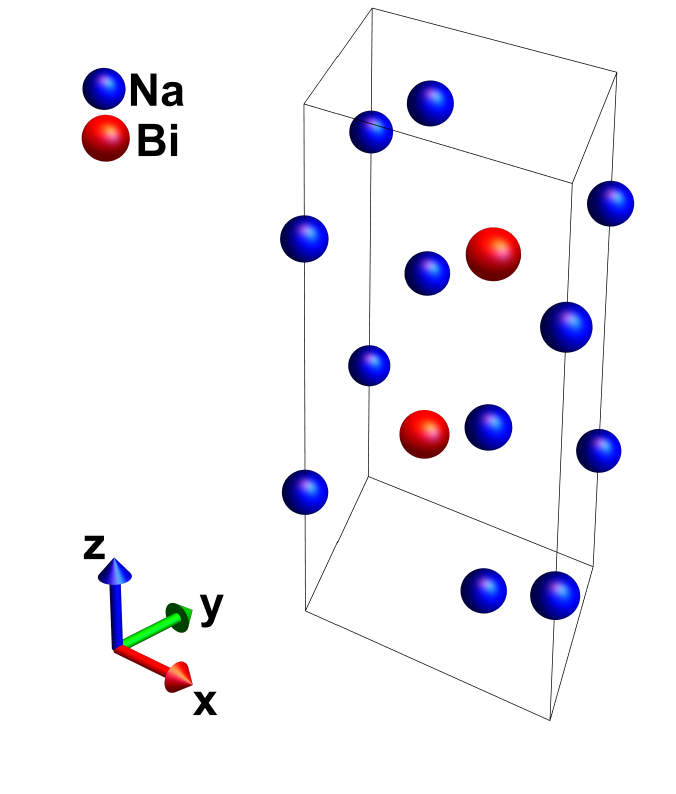}}
	\subfloat[\label{subfig:5b}]{
		\includegraphics[width=0.65\columnwidth]{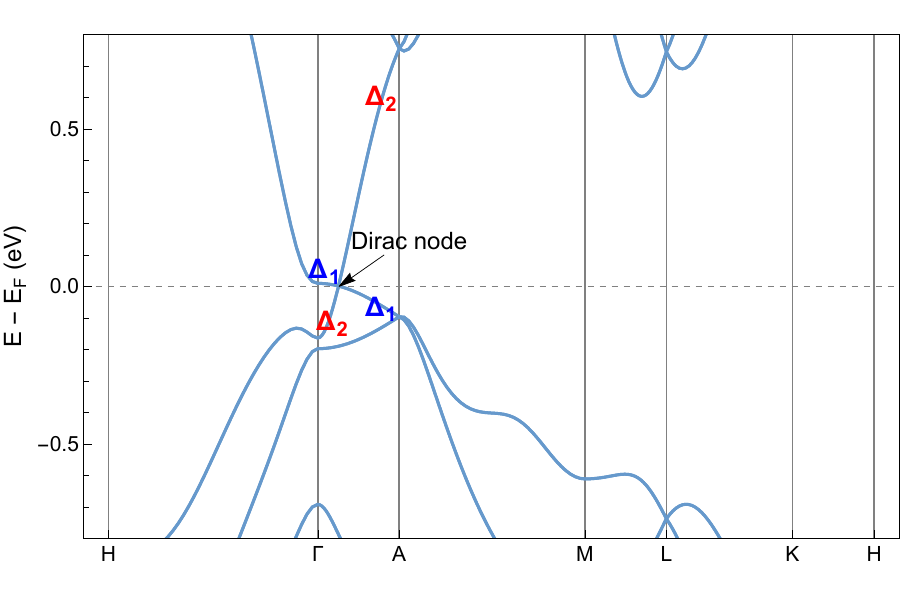}}
	\caption{(a) and (b) display the crystal structure and the energy bands along a closed path in the BZ of Na$_3$Bi, respectively. In panel (b), we also present the corresponding irreps of the energy bands near the Fermi energy along HSL $\Delta: (0,0,w)$. The irreps labeled \(\lbrace 1 \rbrace\) and \(\lbrace 2 \rbrace\) along HSL $\Delta: (0,0,w)$ are represented in blue and red, respectively. The coordinates of the HSPs are as follows: $\text{H}: (0.333333, 0.333333, 0.5)$, $\Gamma : (0,0,0)$, $\text{A}: (0,0,0.5)$, $\text{M}: (0.5,0,0)$, $\text{L} : (0.5, 0, 0.5)$, and $\text{K} : (0.333333, 0.333333, 0)$.}
	\label{fig:5}
\end{figure}

First, we create a new directory, \texttt{workingdir}, and then use the \texttt{Getstruct} function to generate the \texttt{struct.mx} file. For the non-magnetic material Na$_3$Bi, we do not need to modify the \texttt{struct.mx} file. The content of the \texttt{struct.mx} file is as follows in Figure \ref{s3}:

\begin{figure}[htbp]
	\centering
	\includegraphics[width=1\columnwidth]{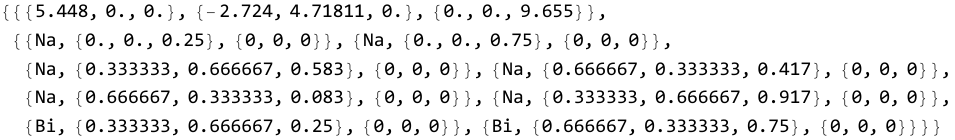}
	\caption{The structure data of Na$_3$Bi in the \texttt{struct.mx} file.}
	\label{s3}
\end{figure}
From Figure \ref{s3}, we can observe that the primitive lattice vectors are $\vec{a}_1=(5.448, 0, 0)$, $\vec{a}_2=(-2.724, 4.718, 0)$, and $\vec{a}_3=(0, 0, 9.655)$, expressed in units of \AA\ within the Cartesian coordinate system $Oxyz$. The Na atoms are located at $(0, 0, 0.25)$, $(0, 0, 0.75)$, $(0.333, 0.667, 0.583)$, $(0.667, 0.333, 0.417)$, $(0.667, 0.333, 0.083)$, and $(0.333, 0.667, 0.917)$, while the Bi atoms occupy $(0.333, 0.667, 0.25)$ and $(0.667, 0.333, 0.75)$, relative to the lattice vectors $\vec{a}_1$, $\vec{a}_2$, and $\vec{a}_3$. The corresponding crystal structure is depicted in Figure \ref{subfig:5a}.

\begin{figure}[htbp]
	\centering
	\includegraphics[width=1\columnwidth]{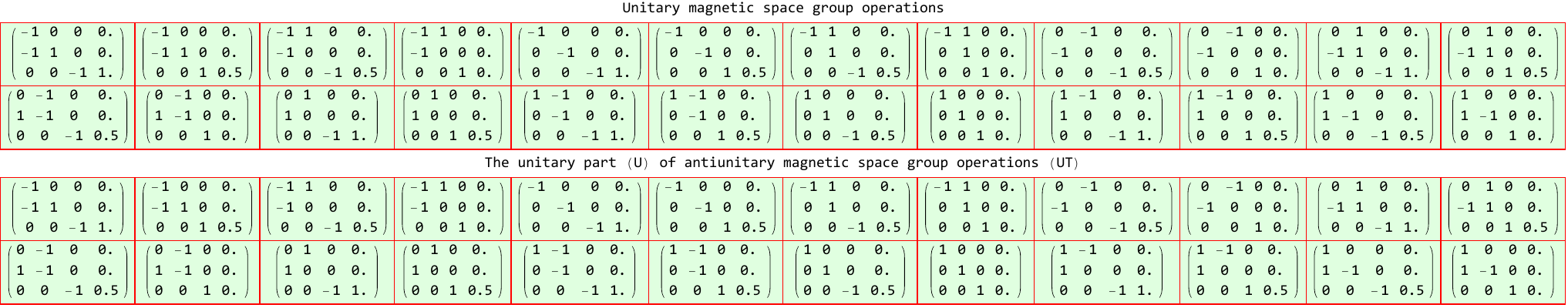}
	\caption{This figure shows the output of \texttt{ShowMSGop[workingdir]}, which lists all the MSG operations for the given Na$_3$Bi structure.
	}
	\label{n1}
\end{figure}

To further investigate the system, we use the command \texttt{Getgkinfo[workingdir, struct, 0.001, 1, 1, 2, 1, 50, 1]} in \texttt{workingdir}, which generates five key files: \texttt{msgop.mx}, \texttt{transformation.mx}, \texttt{gkHSLinfo.mx}, \texttt{gkHSLinfo1.mx}, and the KPOINTS file. The \texttt{ShowMSGop} function is subsequently used to visualize the associated MSG operations, as shown in Figure \ref{n1}.

\begin{figure}[htbp]
	\centering
	\includegraphics[width=0.9\columnwidth]{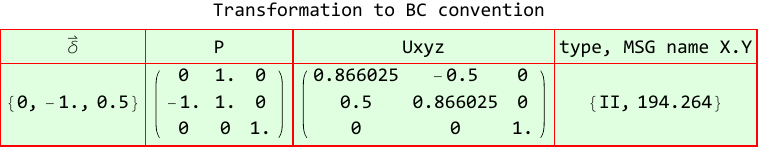}
	\caption{This figure shows the output of \texttt{ShowTrantoBC[workingdir]}, which provides the three parameters $\vec{\delta}$, $P$, and $U_{\text{xyz}}$ mentioned earlier, describing the transformation from the given crystal structure convention of Na$_3$Bi to the BC convention.}
	\label{n2}
\end{figure}

After we talk about the MSG operations, let us give the transformation from the customized crystal structure convention to the BC convention and use the \texttt{ShowTrantoBC} function to visualize this transformation, as shown in Figure \ref{n2}.

\begin{figure}[htbp]
	\centering
	\includegraphics[width=0.9\columnwidth]{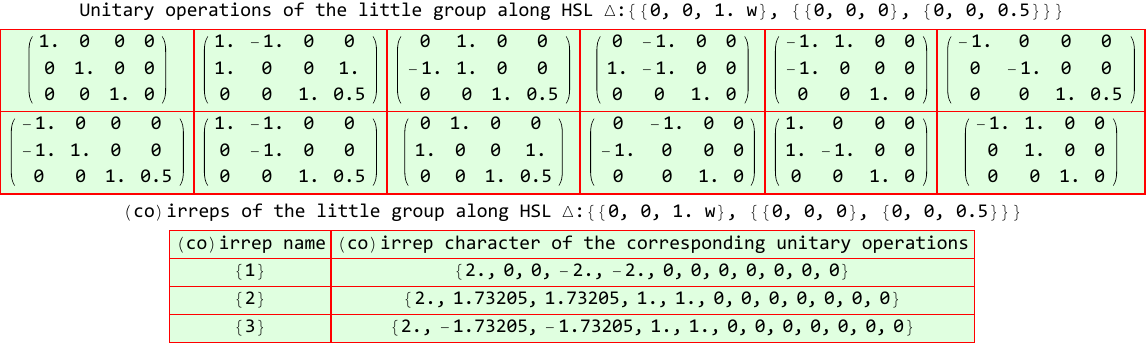}
	\caption{This figure displays the output of \texttt{ShowKpointinfo[workingdir, 2, 1]}. For each unitary MSG operation $g$, we show only the first character of the double-valued representation. The complete results can be found in the \texttt{gkHSPinfo.mx} file in the working directory.}
	\label{n3}
\end{figure}
Using the function \texttt{ShowKpointinfo[workingdir, 2, 1]}, where the first argument "2" refers to the HSL and the second "1" to the first HSL in the \texttt{gkHSLinfo.mx} file within the working directory, we demonstrate the unitary MSG operations of the little group for the $\vec{k}$-points along HSL \(\Delta: (0,0,w)\), as well as the corresponding character table of its (co)irreps, as shown in Figure \ref{n3}. By considering the primitive lattice vector \(\vec{a}_3\), we deduce that HSL \(\Delta: (0,0,w)\) is aligned along the \(k_z\)-axis. Subsequently, we will discuss and predict the presence of a four-degenerate Dirac node on this HSL.

Figure \ref{subfig:5b} shows the energy bands along a closed path in the BZ, along with the (co)irreps of the energy bands near the Fermi energy, specifically along the HSL \(\Delta: (0,0,w)\) of Na$_3$Bi. In Figure \ref{subfig:5b}, the irreps labeled \(\lbrace 1 \rbrace\) and \(\lbrace 2 \rbrace\) along HSL \(\Delta: (0,0,w)\) are represented in blue and red, respectively.

\begin{figure}[H]
	\centering
	\subfloat[\label{subfig:53a}]{
		\includegraphics[width=0.9\columnwidth]{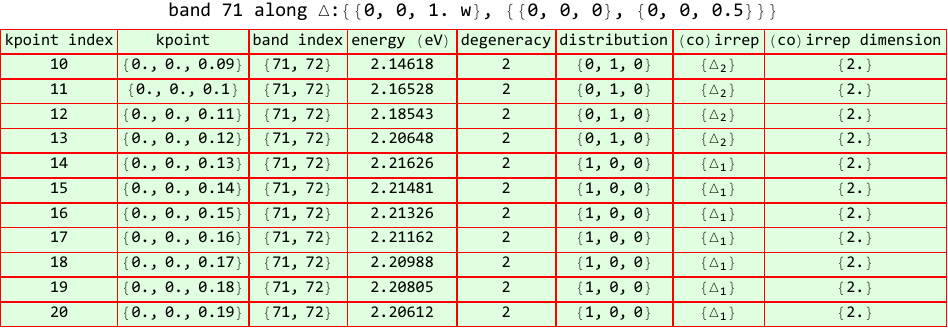}} \\
	\subfloat[\label{subfig:53b}]{
		\includegraphics[width=0.9\columnwidth]{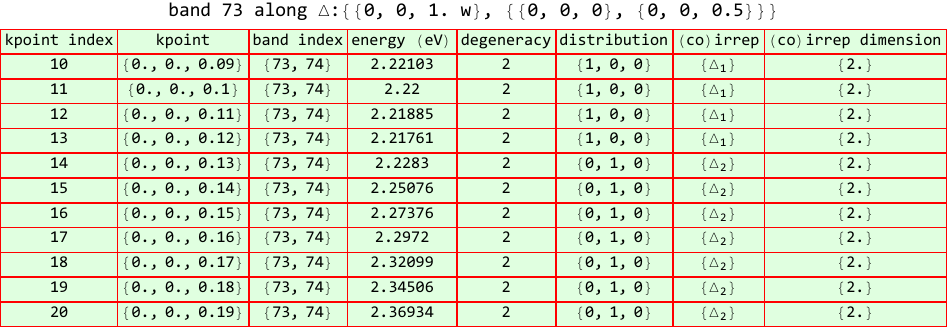}} \\
	\caption{(a) and (b) are the output of \texttt{ShowBandirrepline[workingdir, 2, 1, 71, 10, 20]} and \texttt{ShowBandirrepline[workingdir, 2, 1, 73, 10, 20]} functions, respectively.}
	\label{fig:53}
\end{figure}

From Figure \ref{fig:5}, we can see that along the HSL \(\Delta: (0,0,w)\) from the HSP \(\Gamma\) to the HSP A, the irreps of the energy bands near the Fermi level are inverted. The higher energy level changes from \(\Delta_1 \rightarrow \Delta_2\), while the lower energy level changes from \(\Delta_2 \rightarrow \Delta_1\). To further explore this behavior, like before, we use \texttt{GetKpointTop[workingdir, 0.00001, 0.01, 2, 1, 0, 1]} to generate HSLbandirrep1.mx file. Then we apply the \texttt{ShowBandirrepline[workingdir, 2, 1, 71, 10, 20]} and \texttt{ShowBandirrepline[workingdir, 2, 1, 73, 10, 20]} functions to show the irreps of energy band 71 and energy band 73 along the HSL \(\Delta: (0,0,w)\) near the Fermi level, as illustrated in Figures \ref{subfig:53a} and \ref{subfig:53b}, respectively. The $\vec{k}$-point coordinates range from \((0,0,0.09)\) to \((0,0,0.19)\). It can be observed that as the $\vec{k}$-point coordinates change from \((0,0,0.12)\) to \((0,0,0.13)\), the irrep of the two-degenerate band \(\lbrace 71, 72 \rbrace\) changes from \(\Delta_2\) to \(\Delta_1\), while the irrep of the two-degenerate band \(\lbrace 73, 74 \rbrace\) changes from \(\Delta_1\) to \(\Delta_2\). This indicates the presence of a four-degenerate Dirac node between \((0,0,0.12)\) and \((0,0,0.13)\), which is pointed to by the black arrow in Figure \ref{subfig:5b}.

Finally, we use the function \texttt{GetKpointTop[workingdir, 0.00001, 0.01, 1, 1, 1, 1]} to generate the file \texttt{HSPTop.mx}, which serves as the input file for the package \texttt{CalTopoEvol}. By using \texttt{CalTopoEvol}, we find that there are two Dirac nodal points along HSL \(\Delta: (0,0,w)\). Through the compatibility relations, we can obtain the \(k \cdot p\) model for the Dirac nodal points.

\section{CONCLUSION AND PERSPECTIVE}\label{5}
In this paper, we provide a comprehensive introduction to the functionalities and usage of the Mathematica-based computational package \texttt{ToMSGKpoint}, which we have developed. With just a few simple steps, users can obtain the MSG operations for any crystal structure, the transformation from their own primitive crystal structure convention to the BC convention, little groups of HSPs, HSLs, and HSPLs, their (co)irreps, as well as the (co)irreps of energy bands. To demonstrate the capabilities of the package, we present a detailed computational workflow using nonmagnetic materials Bi$_2$Se$_3$, Na$_3$Bi, and the AFM material MnBi$_2$Te$_4$ as examples.

We would like to emphasize that the above calculations are based on VASP-calculated electronic wavefunctions. For phonons, the little groups and single-valued (co)irreps of the little groups at HSPs, HSLs, and HSPLs can be obtained using our \texttt{ToMSGKpoint} package. By applying the little groups of HSPs, HSLs, and HSPLs to the phonon wavefunctions computed by the \textit{ab initio} package PHONOPY \cite{v3}, we can similarly determine the (co)irreps of the phonon bands. The same approach applies to other bosons, such as magnons and photons. As for the emerging field of spin space groups, following a similar workflow as outlined in this package, we can also compute the little groups of HSPs, HSLs, and HSPLs, and the (co)irreps of the bands under spin space group symmetry in future work.

\section{ACKNOWLEDGMENTS}\label{6}
This paper was supported by the National Natural Science
Foundation of China (NSFC) under Grants No.12188101, No.12322404, No.12104215, No.11834006, the National Key
R\&D Program of China (Grant No.2022YFA1403601), and
Innovation Program for Quantum Science and Technology,
No. 2021ZD0301902. F.T. was also supported by the Young
Elite Scientists Sponsorship Program by the China Association for Science and Technology. X.W. also acknowledges
support from the Tencent Foundation through the XPLORER
PRIZE.


\end{document}